# Tunable Optical Coherence Tomography in the Infrared Range Using Visible Photons

Anna V. Paterova[1,2], Hongzhi Yang[1], Chengwu An[1], Dmitry A. Kalashnikov[1],
and Leonid A. Krivitsky[1,*]

[1] Data Storage Institute, Agency for Science Technology and Research (A*STAR), 138634 Singapore
[2] School of Electrical and Electronic Engineering,
Nanyang Technological University, 639798 Singapore
*E-mail: Leonid-K@dsi.a-star.edu.sg

We report a proof-of-concept demonstration of a tunable infrared (IR) optical coherence tomography (OCT) technique with detection of only visible range photons. Our method is based on the nonclassical interference of frequency correlated photon pairs. The nonlinear crystal, introduced in the Michelson-type interferometer, generates photon pairs with one photon in the visible and another in the IR range. The intensity of detected visible photons depends on the phase and loss of IR photons, which interact with the sample under study. This enables us to perform imaging and characterize sample properties in the IR range by detecting visible photons. The technique possesses broad tunability and yields a fair axial and lateral resolution. The work contributes to the development of versatile 3D imaging and material characterization systems working in a broad range of IR wavelengths, which do not require the use of IR-range equipment.

## 1. Introduction

Optical coherence tomography (OCT) is an appealing technique in bio-imaging [1], medicine [2] and material analysis [3]. The basic OCT setup represents a Michelson interferometer with a sample under test placed in one of the arms (Fig. 1(a)). A probing light beam is split into two arms by a beam splitter, and the beams are recombined after being reflected by a reference mirror and a sample. The interference occurs, when the optical path difference between the two arms is within the coherence length $l_{coh}$ of the light source. By translating the reference mirror, the reflectivity and scattering of a sample are measured at different depths.

The resolution and the signal-to-noise ratio of the OCT are strongly dependent on scattering and absorption of the sample. Most commercially available OCT systems operate in the near-infrared (IR) range, typically in 850-930 nm band, O- (1260-1360 nm) and C- (1530-1565 nm) bands, which is a compromise between scattering and absorption in bio-tissues. Indeed, the Rayleigh scattering is inversely proportional to the wavelength, while the absorption in tissues in near-IR grows with the wavelength. At the same time, for many applications, including analysis of ceramics and polymers, the OCT signal is mainly hampered by the scattering. In this case, OCT measurements in mid- and far- IR lead to significantly more accurate results [4].

Reported mid-IR OCT systems use multiregion quantum cascade lasers and mid-IR photodetectors [5]. More recently high power low-coherence quantum cascade superluminescent emitters have been introduced [6]. In these systems, both the light source and the photodetector require cryogenic cooling.

The development of accessible OCT systems in mid- and far-IR ranges is still a challenging task, and to the best of our knowledge, no commercial system exists yet.

Using nonclassical light sources can be beneficial for the further development of OCT. For example, the method, referred to as a quantum OCT (QOCT), is based on the interference of two indistinguishable photons in the Hong-Ou-Mandel interferometer [7]. The OCT signal is obtained from correlation measurements of photocounts of two detectors. The QOCT allows increasing the resolution by the factor of two compared to classical analogs and eliminates the effect of group velocity dispersion [8, 9].

In this work, we further extend the use of nonclassical light and realize a new method of OCT in which measurements in IR range are performed via detection of visible photons. The wavelength of the probing photon can be tuned to minimize the scattering, while the wavelength of the detected photon can be set in the detection-friendly region. Recently the concept of nonlinear interference was used for to perform 2D imaging [12], interferometry with sensitivity below the shot-noise [13], and IR spectroscopy with visible light [14-17]. We further develop the technique to demonstrate imaging through an opaque substrate, measurements of optical depths and sample birefringence in a broad IR range, all with the same measurement configuration. Our approach is based on a nonlinear Michelson interferometer, which is versatile and straight forward to implement. So far, none of the previously reported schemes have achieved the same level of functionality in a single configuration. Our work contributes to the development of 3D imaging and material characterization systems operating over a broad range of IR wavelengths.

## 2. Method

The scheme of our method is shown in Fig. 1(b). Correlated photons (signal and idler) are generated via frequency non-degenerate collinear spontaneous parametric down conversion (SPDC) in a nonlinear crystal [18, 19]. The phase matching conditions are chosen in such a way, that the wavelengths of signal (detected) and idler (probe) photons are in the visible and IR range, respectively. A dichroic mirror DM separates the photons. Signal and pump photons are reflected by a reference mirror M and the idler photon is reflected by the sample under study. Another pair of photons (identical to the first one) is generated from the second pass of the pump through the crystal.

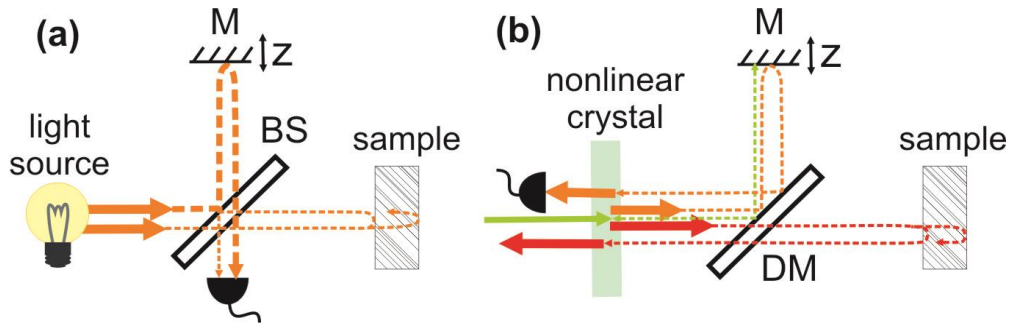

Fig. 1. The scheme of (a) the conventional and (b) the proposed OCT setup. Arrows indicate the interfering photons; dashed lines show the photons paths. In (b) the laser generates signal (visible) and idler (IR) photons in the nonlinear crystal. Photons propagate collinearly and are split by a dichroic mirror (DM). Pump and signal photons are reflected by a mirror (M); idler photons are reflected back by a sample under study. Interference of visible photons is detected as a function of the displacement of mirror M. Properties of the sample in the IR range are inferred from the interference of photons in the visible range. Green, orange, and red arrows stand for the pump, signal and idler photons, respectively. The beams are shifted for clarity.

Once optical paths of signal and idler photons are equalized, the photon pairs generated from the first and the second passes of the pump through the crystal interfere. The interference is associated with the effect of induced coherence without induced emission, first studied by Zou, Wang and Mandel [10, 11]. The explicit theoretical description of the nonlinear Michelson interferometer is given in [16]. In brief, let $|1\rangle_{s,i} = a^+_{s,i}|vac\rangle$ denote the single-photon Fock state, where $a^+_{s,i}$ are photon creation operators for signal (*s*) and idler (*i*) photons, respectively, and $|1\rangle_{i_0}$ indicates lossy idler modes. For a perfectly aligned interferometer the state vector is given by the superposition of two-photon states created at the first and the second pass of the pump through the nonlinear crystal:

$$|\Psi\rangle \propto \left[|\tau_i|^2 e^{i(\varphi_s+\varphi_i)}|1\rangle_s|1\rangle_i + \sqrt{1-|\tau_i|^4}|1\rangle_s|1\rangle_{i_0}\right] + e^{i\varphi_p}|1\rangle_s|1\rangle_i, \quad (1)$$

where $\tau_i$ is the amplitude transmission coefficient of the idler photons for a single pass through the sample, which accounts for all the losses due to reflection, absorption and scattering; $\varphi_{s,i,p}$ are the phases acquired by the signal, idler and pump photons respectively. Now, we explicitly introduce the reflection coefficient of idler photons by the sample $r_i$, which leads to the following transformation:

$$|\tau_i|^2 \rightarrow |\tau_i|^2 |r_i|. \quad (2)$$

To account for a finite coherence length of down-converted photons in Eq. (1) we introduce the normalized first-order correlation function of the SPDC, given by the Fourier transform of the SPDC spectrum:

$$|\mu(\Delta t)| = 2\pi \int_0^\infty d\Omega \, |S(\Omega)| \, e^{-i\Omega \Delta t}, \quad |\mu(0)| = 1, \quad (3)$$

where $\Delta t$ is the time delay between signal and idler photons in the interferometer, $|S(\Omega)|$ is the intensity spectrum of SPDC photons; $\Omega$ is the frequency detuning.

The probability of detecting a signal photon is given by $I_s = \langle \Psi | a^+_s a_s | \Psi \rangle$, and the resulting intensity dependence for the signal photons is given by [8-15, 17, 20, 21]:

$$I_s \propto \left(1 + |r_i||\tau_i|^2 |\mu(\Delta t)| \cos(\varphi_s + \varphi_i - \varphi_p)\right), \quad (4)$$

From Eq. (4) it follows that the interference pattern for signal photons depends on the phase shifts and losses experienced by idler photons. Thus sample properties (transmittance, reflectivity and refractive index) at the wavelength of idler photons can be inferred from measurements of the interference pattern of signal photons. Direct detection of idler photons is not required.

When interferometer arms are aligned ($\Delta t = 0$), the visibility of the interference pattern is given by $V = \left(I_s^{\max} - I_s^{\min}\right) / \left(I_s^{\max} + I_s^{\min}\right) = |r_i||\tau_i|^2$. For the sample with $j \in [1, N]$ reflecting surfaces, the interference signal can be found at several positions of the reference mirror M. Corresponding visibility values are given by:

$$V_j = |r_j(z_j)| \cdot \prod_{m=0}^{j-1} |\tau_m(z)|^2 \cdot \left(1 - |r_m(z_m)|^2\right), \tag{5}$$

where $|r_j(z_j)|$ is the amplitude reflection coefficient from the *j*-th boundary at the depth $z_j$, $\tau_m(z)$ is the amplitude transmission coefficient determined by the absorption and scattering inside the sample, and the last term corresponds to losses due to reflection from previous (*j-1*) surfaces. Thus from the measurements of the interference visibility for signal photons we determine the reflection coefficient of the sample for idler photons and from the separation between the interference signals, we infer the optical thickness of the sample. The axial resolution of our method is defined by the width of the correlation function $|\mu(\Delta t)|$, which depends on the bandwidth of SPDC photons [20].

## 3. Experiment

Our experimental setup is shown in Fig. 2. The pump beam of the continuous wave (CW) laser (Nd:YAG at 532 nm, 80 mW or Ar+ at 488 nm, 34 mW) is reflected by the dichroic mirror $DM_1$. The lens $F_1$ (*f*=200 mm) focuses the pump into the periodically poled Lithium Niobate (PPLN) crystal (length 1 mm), where SPDC occurs. Lenses F' (*f* =75 mm) form collimated pump, signal and idler beams after the crystal [22, 23].

The photons propagate collinearly and are separated into two arms of the interferometer by the dichroic mirror $DM_2$. A silver coated mirror $M_s$, mounted on the motorized translation stage, reflects signal and pump photons. A sample under test is inserted in the path of idler photons. In our experiment we used several samples: a Silicon window (Edmund Optics), a compound retardation waveplate (Thorlabs) and a Chromium coated microscope resolution target test (Negative 1951 USAF microscope target on glass substrate, ThorLabs). In the absence of a sample, the gold-coated mirror $M_i$ is used as a reference.

The reflected pump beam generates another pair of photons in the PPLN crystal. We scan the position of the mirror $M_s$ and observe the modulation of the intensity of signal photons using a Silicon avalanche photodiode for data acquisition (Perkin Elmer, AQR-14FC) or a CCD camera (Thorlabs) for system alignment, see supplementary materials.

We use a single off-shelve PPLN crystal (Covesion) with nine inscribed regions with different poling periods. The wavelength of SPDC photons is chosen via the selection of the poling period ($\Lambda_{PPLN}$) and the crystal temperature ($T_{PPLN}$). For this demonstration we perform experiments at four different wavelengths of idler photons, see Table I:

Table I. Experimental configurations

| Probe photon wavelength (FWHM), nm | Detected photon wavelength (FWHM), nm | Pump wavelength, nm | Coherence length, μm | PPLN temperature, K | PPLN poling period, μm |
|---|---|---|---|---|---|
| 1543 (29) | 812.2 (8.2) | 532 | 80±2 | 399 | 7.4 |
| 2140 (24.5) | 707.9 (2.7) | 532 | 185±3 | 465 | 8.36 |
| 2504 (35) | 606.1 (2.0) | 488 | 184±3 | 399 | 7.4 |
| 3011 (43) | 582.4 (1.61) | 488 | 211±3 | 465 | 8.03 |

The spectra of detected photons are measured using the home-built grating spectrometer, see results in the supplementary material. The coherence length is given by $l_{coh} = \lambda^2/\Delta\lambda$. The increase of the coherence length is expected from the phasematching conditions, which predict the narrowing of the bandwidth of the SPDC spectrum, see supplementary materials.

For raster imaging (A-scan) of the microscope resolution target test, we insert an additional imaging lens F (*f*=50 mm) in the path of idler photons. The focal plane of the lens is positioned on the sample surface. For each position of the sample, we measure the interference visibility of visible photons by scanning the mirror $M_s$. By translating the sample in the plane transverse to the incident beam we reconstruct its lateral image.

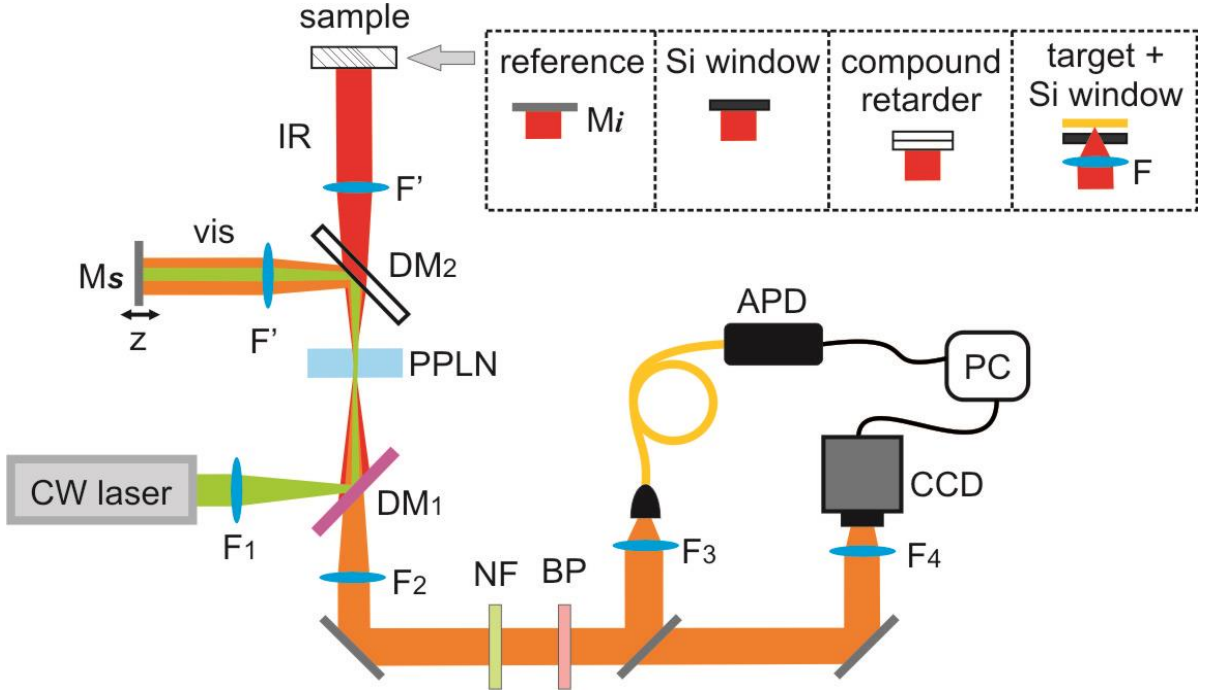

Fig. 2. The experimental setup. A CW laser is used as a pump, and it is injected through the dichroic mirror $DM_1$. SPDC photons are produced collinearly in the PPLN crystal. The signal, pump and idler photons are separated into two channels by the dichroic mirror $DM_2$. Signal (visible) and pump photons propagate together towards the scanning mirror $M_s$. Idler (IR) photons are sent to the sample. Lenses F' collimate the beams. The interference of signal photons is observed as a function of the position of the mirror $M_s$ either by an APD or a CCD camera. Lens $F_2$ collimates the signal beam. Noise is filtered by the dichroic mirror ($DM_1$), notch (NF) and bandpass (BP) filters. The sample is placed into IR arm of the interferometer. Mirror $M_i$ is used for calibration. In the imaging configuration, an additional lens F is introduced into the IR arm of the interferometer.

## 4. Results

### 4.1 Measurements of positions of reflective layers

First, we describe the experiment with probing photons at $\lambda_i$ = 1543 nm (idler) and detected photons at $\lambda_s$ = 812 nm (signal). For system calibration, we measure the intensity modulation of signal photons without the sample. We insert the gold mirror $M_i$ in the IR arm and scan the position of the mirror $M_s$. Once the optical paths of signal and idler photons are equalized we observe the modulation of the intensity of signal photons (Fig. 3(a)). We chose this point as an origin for the mirror displacement (z=0 mm). The visibility of the interference pattern is measured from the fine scan of the mirror $M_s$, and it is equal to $V_{ref} = 81 \pm 1\%$ (Fig. 4(a)). Note that, in accordance with Eq. (4), the modulation period of interference fringes for signal photons is given by the wavelength of idler photons.

Then we insert the sample and repeat the scan (Fig. 3(b)). We use an uncoated pure Silicon window (Edmund Optics) with a thickness of $1080 \pm 4$ μm. The absorption and scattering coefficients for the sample are negligible at the probing wavelengths. Assuming a non-absorptive sample with $N=2$ surfaces, the visibility ratios are given by (see Eq. (5)):

$$\begin{cases} V_1 / V_{ref} = |r_1(z_1)| / |r_{ref}|, & (|\tau_0| \equiv 1, |r_0| \equiv 0) \\ V_2 / V_{ref} = |r_2(z_2)| \cdot (1 - |r_1(z_1)|^2) / |r_{ref}| \end{cases}, \quad (6)$$

where $|r_1(z_1)|$, $|r_2(z_2)|$ are the amplitude reflection coefficients by front and back surfaces of the mirror respectively, and $|r_{ref}|$ is the amplitude reflection coefficients of the reference mirror $M_i$.

The detected interference signals at $z_1 = -10.1$ mm and $z_2 = -6.2$ mm correspond to the reflection of idler photons from the front and back surface of the Silicon window, respectively. The interference observed at $z_{12} = -2.3$ mm corresponds to the second reflection of idler photons from two surfaces of the Silicon window. Its visibility is equal to $V_{12} = V_2 \cdot |r_1(z_1)| \cdot |r_2(z_2)|$. The visibility for the signal reflected from the back surface and multiple reflections is less than for the signal reflected from the front surface due to the loss of idler photons, see Eq. (6).

Fig. 3(c, e, g) show measurement results with the reference mirror $M_i$ with probing idler photons at 2140 nm (detected signal at 708 nm), 2504 nm (detected signal at 606 nm), and 3011 nm (detected signal at 582 nm) respectively. The visibility of the interference is measured from the fine scan of the mirror $M_s$, see Fig. 4(b, c. d). It is equal to $33 \pm 1\%$ at 2140 nm, $20 \pm 1\%$ at 2504 nm and $20 \pm 1\%$ at 3011 nm. The visibility in Figs. 4(b, c, d) is somewhat lower than in Fig. 4(a), mainly due to imperfections of the dichroic mirror $DM_2$, as confirmed by independent measurements.

Experimental results with the Silicon sample with probing photons at 2140 nm, 2504 nm and 3011 nm are shown in Fig. 3(d, f, h), respectively. Similar to the measurement at 1543 nm, the interferograms show reflections from the front, back and multiple surfaces of the sample. As expected, the visibility of signal reflected from the back and multiple surfaces is decreased due to the gradual loss of idler photons.

From Eq. (6) we calculate reflection coefficients of the Silicon at all the four probing wavelengths, see Table II. Corresponding refractive indices are calculated using Fresnel equations. Our results are in good agreement with the database values [24]. As expected, the refractive index for Silicon is decreasing with the increase of the wavelength of probing photons. The achieved accuracy of the reflectance measurements is $\Delta R_i = \pm 0.003$ and of the refractive index is $\Delta n_i = \pm 0.014$.

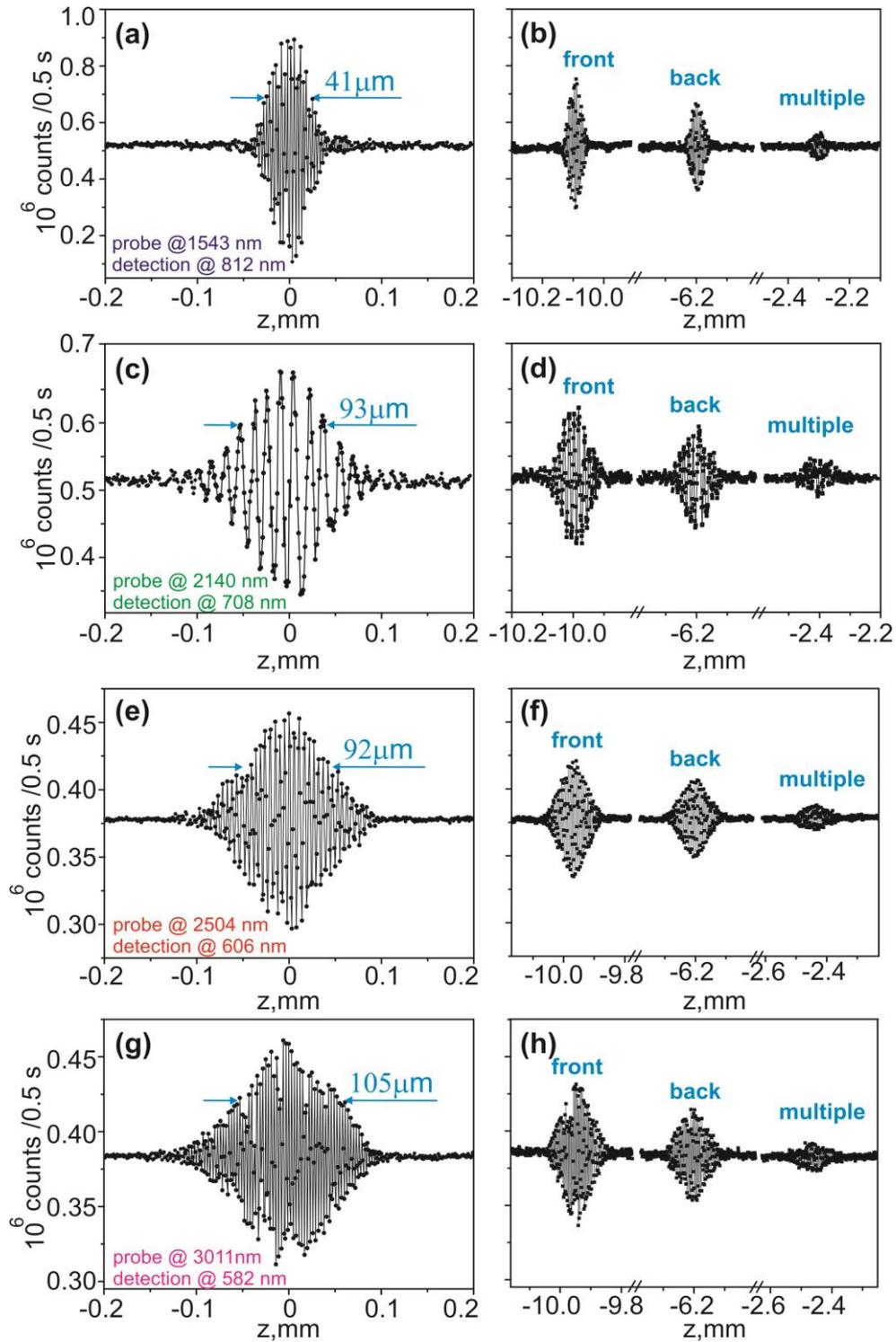

Fig. 3. The intensity of signal photons detected by the APD versus the position of the mirror $M_s$. Results in (a) and (b) are obtained for probing photons at $\lambda_i$=1543 nm (detected photons at $\lambda_s$=812 nm); in (c) and (d) at $\lambda_i$=2140 nm ($\lambda_s$=708 nm); in (e) and (f) at $\lambda_i$=2504 nm ($\lambda_s$=606 nm); in (g) and (h) at $\lambda_i$=3011 nm ($\lambda_s$=582 nm). (a), (c), (e) and (g) are signals obtained with the reference mirror $M_i$; (b), (d), (f) and (h) are signals obtained with the Silicon window sample when IR photons are reflected from different surfaces of the sample.

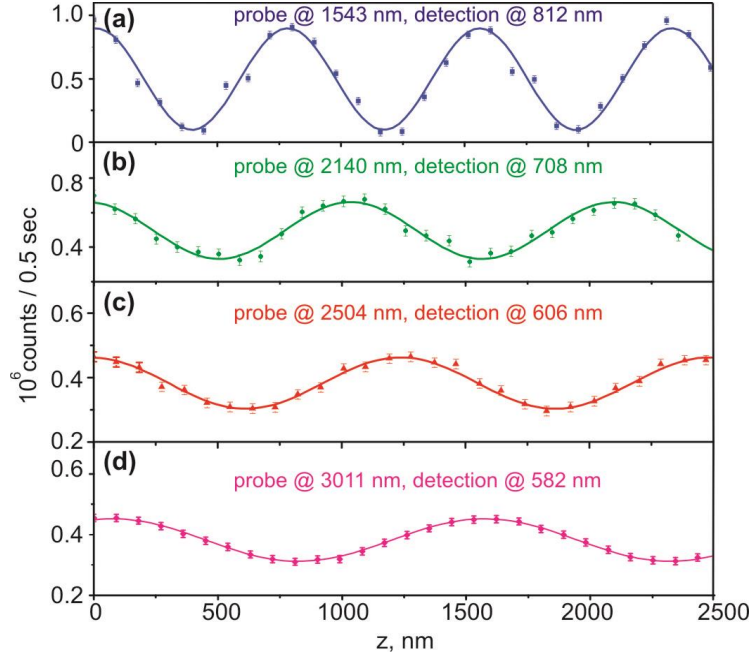

Fig. 4. Fine scans of interference fringes shown in Fig. 3(a, c, e, g). Dependence of intensity of the signal photons on the position of the mirror $M_s$ are obtained for probing photons at (a) $\lambda_i$=1543 nm ($\lambda_s$=812 nm); (b) $\lambda_i$=2140 nm ($\lambda_s$=708 nm); (c) $\lambda_i$=2504 nm ($\lambda_s$=606 nm); (d) $\lambda_i$=3011 nm ($\lambda_s$=582 nm). Points indicate the experimental data; solid curves show the cosine fitting of the data according to Eq. (4). Note that modulation period is given by the wavelength of idler photons.

Table II. Calculated reflectance and refractive indices of the Silicon window

| $\lambda_i$, nm | Measured reflectance, $R_i=|r_i|^2$ | Database reflectance, $R_i$ [24] | Measured refractive index, $n_i$ | Database refractive index, $n_i$ [24] |
|---|---|---|---|---|
| 1543 | 0.306 ± 0.003 | 0.306 | 3.476 ± 0.014 | 3.478 |
| 2140 | 0.302 ± 0.010 | 0.303 | 3.441 ± 0.025 | 3.448 |
| 2504 | 0.287 ± 0.015 | 0.302 | 3.31 ± 0.16 | 3.440 |
| 3011 | 0.286 ± 0.015 | 0.0301 | 3.30 ± 0.16 | 3.433 |

## 4.2 Measurement of a birefringent sample with hidden features

To test our technique in a more sophisticated sample, we carried out measurements with a compound zero-order retardation waveplate (Thorlabs). The waveplate consists of two quartz plates (thickness 934 ± 9 μm and 953 ± 9 μm) with orthogonal axes and an air gap between the plates (thickness 123 ± 9 μm). It is designed to introduce a half wavelength retardance at 532 nm. The waveplate is inserted into the idler arm of the interferometer with its fast optical axis parallel to the polarization of idler photons.

The interference pattern without the sample is shown in Fig. 3(a) and 4(a) (the probing wavelength is 1543 nm, and detected wavelength is 812 nm). The interference signal with the waveplate is shown in Fig. 5. Signals at $z_1$ = -7.16 mm and $z_4$ = -4.11 mm correspond to reflections from the front and back surface of the waveplate, respectively. Two additional peaks at $z_2$ = -5.72 mm and $z_3$ = -5.60 mm correspond to reflections from internal interfaces of the compound waveplate.

From the optical delay between reflections #1 and #2 ("front"-"internal 1") we calculate the refractive index along the fast axis ($n_o$). From the optical delay between reflections #3 and #4 ("internal 2"-"back") we calculate the refractive index along the slow axis ($n_e$). Calculated values are shown in Table III and they are in a good agreement with database values [24].

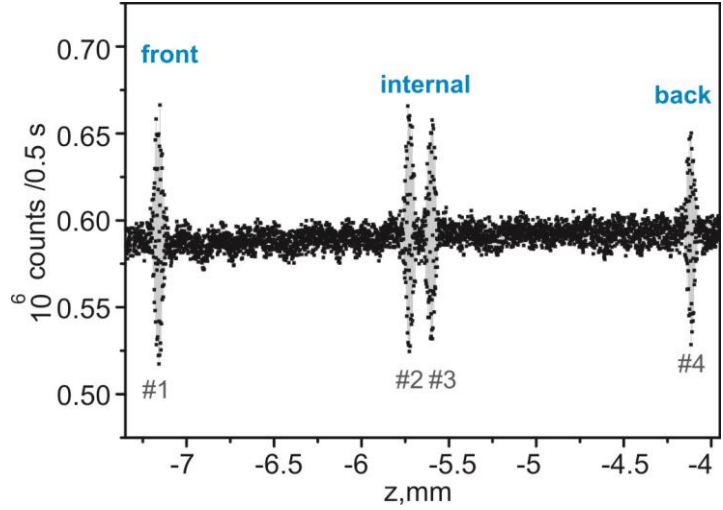

Fig. 5. The intensity of the signal photons detected by the APD versus the position of the mirror $M_s$ with the inserted compound waveplate. Idler (IR) photons are reflected from multiple surfaces of the waveplate. Four interference signals correspond to reflections from the front (#1), two internal (#2, and #3) and the back interfaces (#4). An optical delay between reflections #1 and #2 gives the fast refractive index ($n_o$) and the optical delay between reflections #3 and #4 gives the slow refractive index ($n_e$).

Table III. Calculated reflectance and refractive indices of the compound quartz waveplate.

| $\lambda_i$, nm | Measured refractive index, $n_o$ | Database refractive index, $n_o$ [24] | Measured refractive index, $n_e$ | Database refractive index, $n_e$ [24] |
|---|---|---|---|---|
| 1543 | 1.56±0.04 | 1.528 | 1.57±0.04 | 1.536 |

## 4.3 Reflectance imaging of a sample through a Silicon layer

To show the imaging capability of our technique, we inserted a lens in the path of IR photons and carry out raster imaging (A-scan) of the negative microscope test slide preceded by the AR-coated Silicon window, see Fig 6. The probing wavelength is $\lambda_i$=1543 nm, and detected wavelength is $\lambda_s$=812 nm. IR photons are reflected by different regions of the sample, and we measure the interference patterns for visible photons.

From a raster scan of the sample, we reconstruct its reflectance pattern for IR photons, see Fig. 6(a, b). As expected, the measured reflectivity of the Chromium film at 1550 nm is about 63% [24]. Once the probe beam hits the region without Chromium coating (number "4" in Fig.6(a) and vertical stripes in Fig.6 (b)), we observe almost zero reflectivity. The width of the stripes of the "4" character in Fig.6 (a) is 60 μm. The reflectivity doesn't drop rapidly, as the probing beam size of 50 μm is comparable to the feature size. The width of the test stripes in Fig.6 (b) is 88 μm, and they are well resolved. Similar results have been obtained for other wavelengths of probing photons, see Supplementary materials.

The image of the sample under the white light microscope is shown in Fig. 6(c). Dashed squares show the measured regions of the resolution test slide. The sample cannot be seen under visible light, once the Silicon window is inserted in front of it, see Fig. 6(d). The greenish color appears due to the AR coating of the Silicon window.

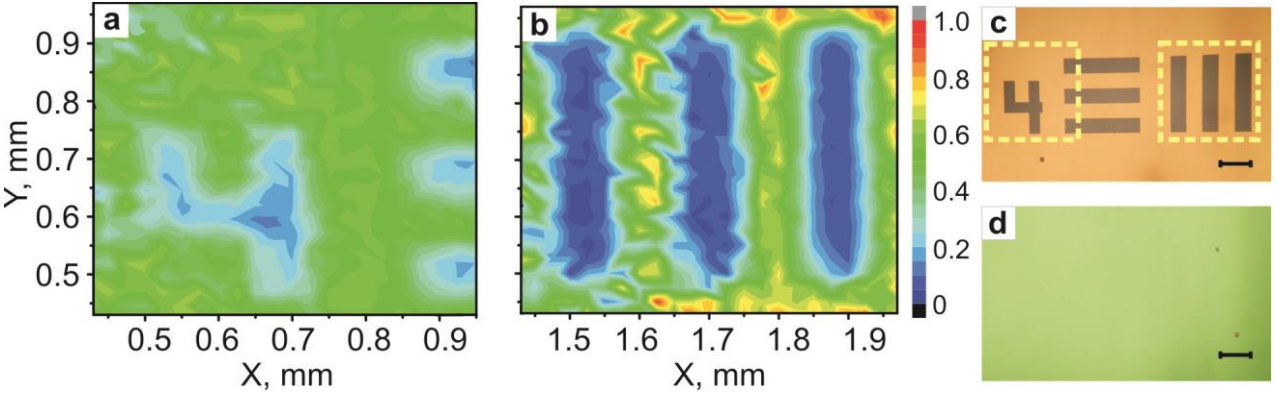

Fig. 6. (a),(b) IR imaging of the microscope test slide preceded by a Silicon window using our technique. High reflectance regions (green) correspond to chrome coated regions; low reflectance regions (blue) correspond to glass substrate. (c), and (d) are white light microscope images of the sample without and with the Silicon window, respectively. Dashed rectangles in (c) highlight measured regions. The scale bar in (c, d) is 200μm.

## 5. Discussion

The axial resolution of the technique is defined by the first order correlation function of the SPDC, given by the Fourier transform of the SPDC spectrum, see Eq. (3). The latter is given by the $sinc^2$-function [18-20]. As a result, interference patterns in Fig. 3 have triangular shapes [18-20]. Even though the demonstrated resolution is on the order of tens of microns, it can be reduced by at least one order of magnitude using dedicated methods of generation of broadband SPDC. The methods include but are not limited to using SPDC crystals with chirped poling period [25], inhomogeneous heating [26] and applying electric field [27] to the SPDC crystal. Moreover, the smooth spectral shape of SPDC is beneficial for OCT in comparison with some classical broadband sources, which is essential for providing high-quality interferograms and better signal to noise ratio [28]. The spatial resolution of the imaging in the lateral plane can be further improved by using focusing lenses with short focal distances or microscope objective lenses.

The central wavelength of the probing beam can be set to the required value by an appropriate choice of the nonlinear crystal and the wavelength of the pumping laser. The range of the operating wavelengths of the method is limited by the absorption of the SPDC crystal in the IR range (up to ~4.5 μm for Lithium Niobate). It can be readily extended into the mid- and far-IR ranges by changing the SPDC crystal and optics. For instance, by using AGS crystal, it is possible to reach the wavelength of idler photons of up to 12 μm.

Sensitivity to the refractive index change is defined by the accuracy of the measurement of the interference visibility. From the results in Table II we calculate that the technique is capable of resolving the refractive index change at the air/sample interface of $\Delta n_i$=0.01. Higher sensitivity can be archived by careful optimization of optical losses and beams overlap.

We note that some of the earlier experiments on 2D imaging and optical sectoring used the nonlinear Mach-Zehnder interferometer [10, 12, 29, 30]. The scheme based on a Michelson interferometer, presented in our work, is much more straightforward and versatile. Indeed, it uses only one nonlinear crystal, which is easier to align and tune the wavelengths. It also requires less optical elements such as lenses and mirrors.

## 6. Conclusions

In summary, we demonstrated the proof-of-concept of tunable at the IR OCT technique, where information about the sample properties at the IR range is inferred from measurements of visible photons. We demonstrate the operation of the technique at four wavelengths: 1543 nm, 2140 nm, 2504 nm, and 3011 nm, with the actual detection in 812 nm, 708 nm 606 nm and 582 nm respectively. The tuning between the wavelengths is rather seamless and requires only the selection of the poling period and the crystal temperature.

The method shows good precision in determining the position of reflecting layers, refractive indices of the sample and its birefringence. In the imaging configuration, the technique allows reflective imaging through a Silicon layer. The resolution and the tuning range can be further improved by a judicious choice of optical elements, such as nonlinear crystals, beam splitters, lenses, etc. Further study of the impact of scattering can help in extending operation of the proposed OCT technique to the mid- and far-IR ranges.

Practical applications of the technique include, but not limited to analysis of highly scattering materials in mid-IR for defectoscopy and photonics industry. Imaging through Silicon layers and thickness analysis is highly relevant to the microelectronics industry as it enables using accessible photodetectors and light sources for imaging through Silicon wafers.

## 7. Acknowledgements

The authors thank D. H. Zhang, R. Bakker and S. Kulik for stimulating discussions. A. P. would like to acknowledge the support of the SINGA PhD fellowship. The work is supported by DSI core funds.

**Supplementary material**

**1. Spectra of detected photons**

Fig. S1 shows spectra of signal photons measured in the experiment. Measurements are taken using a home built grating spectrometer with 0.4 nm resolution.

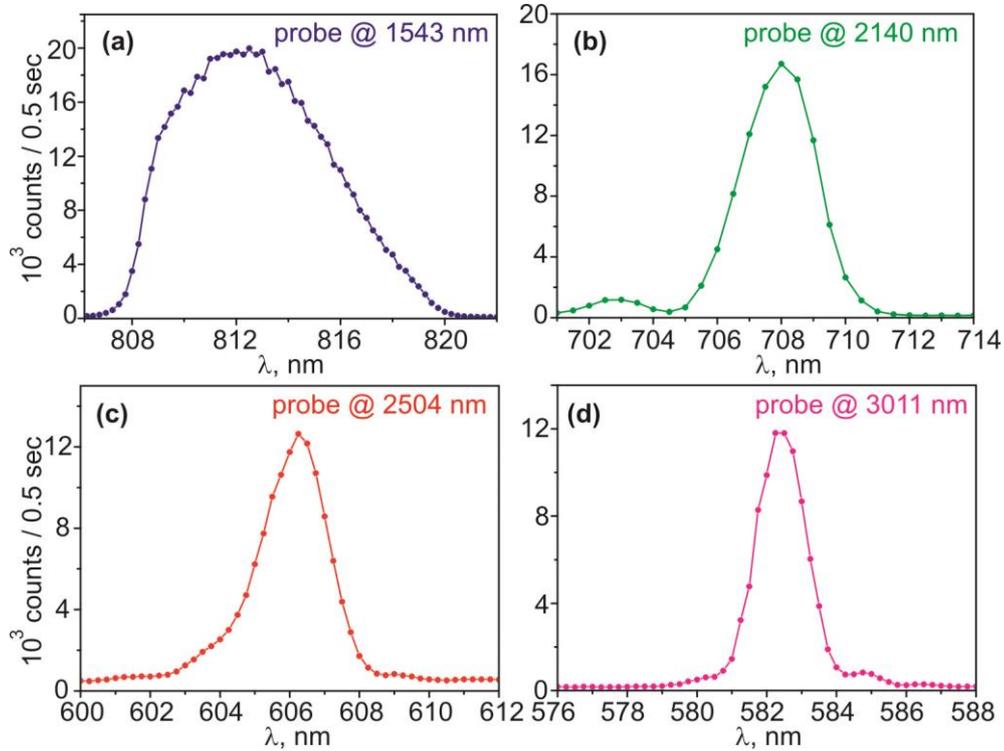

Fig. S1. Spectrum of the detected (signal) photons, when the corresponding probing wavelength is (a) 1543 nm, (b) 2140 nm and (c) 2504 nm, (d) 3011 nm.

**2. Observing interference fringes by a CCD camera**

Fig. S2 shows the interference pattern observed by the CCD camera. The image in Fig. S2(a) corresponds to the reflection from the reference mirror Mi. The interference pattern changes, when a reflection from the front surface of the waveplate is observed (see Fig. S2(b)). The visibility of the interference pattern decreases, since the air/quartz interface reflects only about 3%.

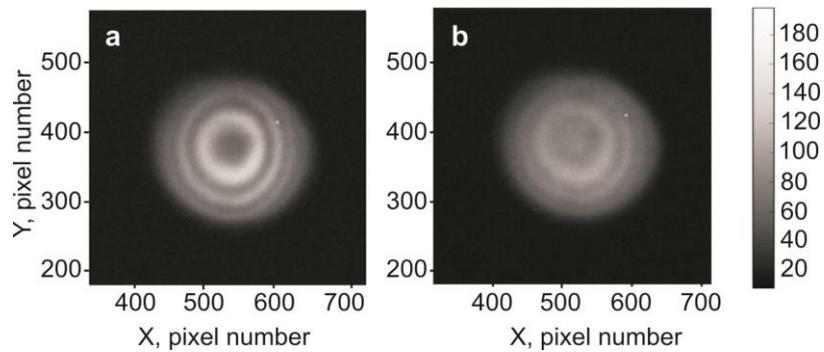

Fig. S2. Interference patterns captured by the CCD camera. (a) Reflection from mirror Mi without the inserted sample. (b) Reflection from the front surface of the waveplate sample.

## 3. Reflectance imaging of a sample through a Silicon layer at $\lambda_i$=2140 nm

We also carried out imaging experiments (A-scan), similar to the one described in Section 4.3 of the manuscript at different wavelengths of probing photons. Fig. S3 shows the image of the region of the microscope test slide (see dashed region in Fig.6 (c)) preceded by a Silicon window, obtained with the wavelength of probing photons at $\lambda_i$=2140 nm, and the wavelength of detected photons at $\lambda_s$=708 nm. The result is consistent with the results shown in Fig.6 of the manuscript.

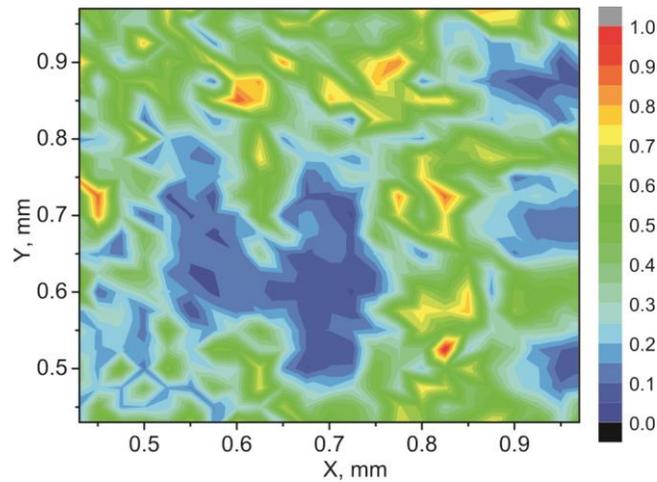

Fig. S3 Infrared imaging of the microscope test slide preceded by a Silicon window using our technique. The wavelength of probing photons is $\lambda_i$=2140 nm, and the wavelength of detected photons is $\lambda_s$=708 nm. High reflectance regions (green) correspond to chrome coated regions; low reflectance regions (blue) correspond to a glass substrate. The imaged region is indicated in by dashed rectangle in Fig. 6(c).